\documentclass[conference]{IEEEtran}
\IEEEoverridecommandlockouts
\usepackage{cite}
\usepackage{amsthm,amsmath,amssymb}
\usepackage{bm}
\newtheorem{definition}{Definition}
\usepackage{graphicx}
\usepackage{textcomp}
\usepackage{xcolor}
\usepackage{subfigure}
\usepackage{bbding}
\usepackage{multirow}

\usepackage{ulem}
\usepackage{flushend}
\usepackage{balance}
\usepackage{color}
\usepackage{booktabs}
\usepackage{wrapfig}

\def\BibTeX{{\rm B\kern-.05em{\sc i\kern-.025em b}\kern-.08em
    T\kern-.1667em\lower.7ex\hbox{E}\kern-.125emX}}
\begin{document}

\title{HierCas: Hierarchical Temporal Graph Attention Networks for Popularity Prediction in Information Cascades
}
\author{
	\IEEEauthorblockN{
		Zhizhen Zhang\IEEEauthorrefmark{2}, 
		Xiaohui Xie\IEEEauthorrefmark{2}$^*$, 
		Yishuo Zhang\IEEEauthorrefmark{4}, 
		Lanshan Zhang\IEEEauthorrefmark{4} 
		and Yong Jiang\IEEEauthorrefmark{2}} 
	\IEEEauthorblockA{\IEEEauthorrefmark{2}Tsinghua University, Beijing, China\\ Email: zhangzz21@mails.tsinghua.edu.cn, xiexiaohui@tsinghua.edu.cn, jiangy@sz.tsinghua.edu.cn}
 
	\IEEEauthorblockA{\IEEEauthorrefmark{3}Beijing University of Posts and Telecommunications, Beijing, China\\ Email: ZhangYishuo1996@outlook.com, zls326@bupt.edu.cn} 

}


\maketitle

\begin{abstract}
Information cascade popularity prediction is critical for many applications, including but not limited to identifying fake news and accurate recommendations. Traditional feature-based methods heavily rely on handcrafted features, which are domain-specific and lack generalizability to new domains. To address this problem, researchers have turned to neural network-based approaches. However, most existing methods follow a sampling-based modeling approach, potentially losing continuous dynamic information that emerges during the information diffusion process. In this paper, we propose Hierarchical Temporal Graph Attention Networks for cascade popularity prediction (HierCas), which operates on the entire cascade graph by a dynamic graph modeling approach. By leveraging time-aware node embedding, graph attention mechanisms, and hierarchical pooling structures, HierCas effectively captures the popularity trend implicit in the complex cascade. Extensive experiments conducted on two real‐world datasets in different scenarios demonstrate that our HierCas significantly outperforms the state‐of‐the‐art approaches. We have released our code at https://github.com/Daisy-zzz/HierCas.
\end{abstract}

\renewcommand{\thefootnote}{\fnsymbol{footnote}}
\footnotetext[1]{Xiaohui Xie is the corresponding author.}
\renewcommand{\thefootnote}{\arabic{footnote}}
\section{Introduction}
In recent years, with the continuous development of social platforms such as Twitter, Weibo and Facebook, people are increasingly using social software to share their lives and opinions. With the development of diverse sharing mediums and sophisticated recommendation algorithms, the speed and breadth of information diffusion have increased significantly\cite{dow2013anatomy,AlFalahi}. Information cascade is a phenomenon that occurs in the process of information spreads between users, e.g., following and retweeting. The involvement of any single person in the middle of the cascade may eventually lead to a dramatic change in the final popularity of the message, which poses a great challenge to cascade popularity prediction. Understanding information cascades becomes important and can lead to significant economic and societal impacts\cite{doi:10.1126/science.aao2998,9078874}.  \par
The cascade of each tweet is usually represented as a graph data structure. The nodes are the set of all re/tweeting users, and the edges represent all retweeting relationships between users in this cascade. Existing works of cascade popularity prediction can be categorized into classification and regression tasks. Regression tasks focus on predicting the precise future popularity value, which can be approached at micro-level (individual user statuses) \cite{deepinf, 10.5555/3367471.3367602} or macro-level (overall cascade outcome) \cite{li2017deepcas, cao2017deephawkes,chen2019cascn}, while the classification tasks aim to predict whether final popularity will be more than a threshold or not. In this study, we focus on macro-level regression tasks, specifically cascade size prediction, i.e. predicting the number of retweets that a post will receive in the future. \par
Traditional feature-based methods heavily rely on handcrafted features, which lack generalizability to new domains\cite{PPOC,10.1145/2433396.2433443,10.1145/2566486.2567997,Shulman2016PredictabilityOP,10.1145/2806416.2806505,10.1145/2661829.2662055}. To address this problem, researchers have turned to neural network-based approaches. Some studies model cascades as a series of propagation paths by using sampling methods such as DeepWalk\cite{10.1145/2623330.2623732}, utilizing recurrent neural networks (RNNs) and attention mechanisms to capture the temporal characteristics of the paths\cite{cao2017deephawkes,li2017deepcas,wang2017topological,10.5555/3172077.3172305,tang2021fully,CasHAN}. However, these approaches often overlook the important effect of topological structure features in cascade dynamics. Inspired by the advancements of graph neural networks (GNNs), some researchers approach the problem by modeling cascades as temporal sequences composed of sub-cascade graphs\cite{chen2019cascn,wang2020predicting,casseqgcn,mucas,casflow}. These methods aim to extract structural information from sub-cascades by employing GNNs. Then RNNs are then utilized to capture the temporal correlation of the sub-cascade sequence. These approaches combine the strengths of both structural and sequential modeling, allowing for a more comprehensive understanding of cascade dynamics. Previous approaches in cascade popularity prediction provide a systematic and structured framework for modeling cascade graphs. Their methods usually consist of three distinct steps, i.e. sampling cascade graphs, extracting structural features, and modeling temporal dependencies. However, by sampling the cascade graph as discrete paths or sub-cascades, the continuous dynamic information that emerges during the information diffusion process is lost. \par
To solve the aforementioned problems, we present Hierarchical Temporal Graph Attention Networks (HierCas). Specifically, HierCas is a layer-wise architecture that stacks several temporal graph attention layers to model the message-passing process on the cascade graph. During the message-passing process, we incorporate time-aware information into the node representation, including relative time and cascade size changes. The relative time is the time interval of the target and neighbors, and the cascade size changes represent the number of new retweets during the time span. Combining size change and relative time information enables the model to understand the dynamic pattern of the cascade graph, i.e. the growth rate of the cascade graph at different moments, so as to predict the trend of future popularity increments. Furthermore, in order to achieve more accurate graph-level predictions, we use a multi-level graph pooling method to dynamically aggregate multiple temporal and structural patterns captured by different graph attention layers.\par
Our main contributions can be summarized as follows:
\begin{itemize}
    \item We propose a novel non-sampling information cascade learning framework HierCas, which jointly learns temporal and structural information on an entire cascade graph using dynamic graph modeling, enabling the capture of popularity trends implicit in complex cascades.
    \item We adopt time-aware node embedding and temporal graph attention layers to capture the temporal dynamics of local structures and global trends in popularity. A multi-level graph pooling method is also utilized to adaptively aggregate different localities for more precise graph-level prediction.
    \item We conduct extensive experiments on several large-scale real-world datasets from two different scenarios, i.e. retweet and citation, demonstrating that HierCas achieves a prediction performance beyond the state-of-the-art approaches.
\end{itemize}

\section{Related Works}
We first introduce some RNN-based methods that model the cascade graph as several propagation paths. DeepCas\cite{li2017deepcas} introduced the first end-to-end deep learning approach for cascade popularity prediction. It samples node sequences using random walks on the cascade graph. These sequences are then processed by a bidirectional GRU (Bi-GRU) with a specially designed attention mechanism to assemble the graph representation. DeepHawkes\cite{cao2017deephawkes} proposed a combination of deep learning and Hawkes process to achieve both high predictive power and interpretability. DeepHawkes captures three important interpretable concepts of Hawkes process: user influence, self-exciting mechanism, and time decay effect in information diffusion. CasHAN\cite{CasHAN} utilizes a node-level attention mechanism based on user influence and a sequence-level attention mechanism based on community redundancy to enhance the representation of the node sequences sampled in the sub-cascade graphs. \par

Considering that the propagation paths will lose the structural information of the cascaded graph, some works sample the cascaded graph as sub-cascade graphs according to the timestamp, and use the graph neural network (GNN)-based model for learning. CasCN\cite{chen2019cascn} introduced the first GNN-based model for popularity prediction. It leverages Graph Convolutional Network (GCN) to capture structural information from the sub-cascade graphs and LSTM to learn temporal dependency between the sub-cascades. Cascade2vec\cite{huang2019cascade2vec} improves the convolutional kernel in the CasCN model and models cascades as dynamic graphs, learning cascade representations via graph recurrent neural networks.  
CasGAT\cite{wang2020predicting} first introduces the use of Graph Attention Networks (GAT) to focus on the local relationship between nodes and their neighborhoods instead of the entire graph. 
VaCas\cite{zhou2020variational} utilizes graph signal processing to model structural information from sub-cascades and Bi-GRU to capture temporal dynamics. The variational autoencoder is first used to model uncertainties in information diffusion and cascade size growth. 
MUCas\cite{mucas} uses GCN to capture structural information from sub-cascade graphs, adopting a dynamic routing mechanism and attention mechanism to learn latent representations of the cascade.
CasHAN\cite{CasHAN} enhances the representation of the node sequences sampled in the cascade graph by a node-level attention mechanism based on user influence and a sequence-level attention mechanism based on community redundancy. 
CasGCN\cite{xu2020casgcn} adopts GCN to learn the entire cascade graph. However, it directly uses timestamps as features and does not utilize parameters for learning, in addition, it uses a static graph modeling approach and loses the dynamic information of cascade graphs \par

From previous works, we can find that most of them design their model following the pipeline: (1) Sample the observed cascade graphs as several sub-cascade graphs. (2) Extract structural features of the sub-cascade graphs with sequential or graph representation methods. (3) Adopt recurrent networks to model temporal dependencies between sub-cascade graphs and get the feature of the entire cascade graph. This pipeline samples the cascade graph as discrete sub-cascades, potentially losing some of the continuous dynamic information that arises during the diffusion of information. As we have noticed recently, during the revision of this work, several papers have been published on the topic of applying dynamic graphs for cascade prediction \cite{lu2023continuoustime,10.1145/3580305.3599281}, which has in fact verified the motivation and significance of our work. However, all of them have overlooked the importance of changes in the cascade size and the summarization of graph-level cascade information for accurately predicting popularity, which have made their contributions
rather orthogonal to ours.

\section{Preliminaries}
Consider a user $u$ posts a message $m$ (e.g., tweets, academic papers, or other information) at time $t_0$, later, other users can interact with this message, e.g., retweets or citations. Given an observation time $T_o$, we assume there are $N$ involved users in total who retweet message $m$ for $K$ times, in result a retweet cascade $C\left(T_o\right)=\left\{\left(u_i, u_j, t_k\right)\right\}_{i, j \in N, k \in K}$, where each 3-tuple represents user $u_i$ retweet user $u_j$'s message at time $t_k \leq T_o$.
\begin{definition}[Cascade Graph] 
    Given a message $m$ and its corresponding retweet cascade $C(T_o)$ before observation time $T_o$, a cascade graph is defined as $\mathcal{G}(T_o)=\left(\mathcal{V}(T_o), \mathcal{E}(T_o)\right)$, with nodes $\mathcal{V}(T_o)=\left\{v_i \mid 0 < i \leq N \right\}$ being the set of all re/tweeting users, and $\mathcal{E}(T_o) \subseteq \mathcal{V}(T_o) \times \mathcal{V}(T_o)$ is the set of directed edges representing all retweeting relationships between users in this cascade.
\end{definition}

From the definition, we can find that $\mathcal{G}(t)$, \text{where} $0 \leq t < T_o$, is time-evolving, and $\mathcal{G}(T_o)$ saves all engaged users and new retweets up to the observation time. The problem definition based on the cascade graph is as follows:

\begin{definition}[Cascade Popularity Prediction] 
    Given a cascade $C(T_o)$ and its cascade graph $\mathcal{G}(T_o)$ observed within the time window $\left[0, T_o\right)$, our goal is to predict the difference between the final popularity $P(T_p)$ and the observed popularity $P(T_o)$ at the initial observation time $T_o$, where $T_p \gg T_o$:
    \begin{equation}
        \Delta P = |\mathcal{C}(T_p)| - |\mathcal{C}(T_o)| = |\mathcal{E}(T_p)| - |\mathcal{E}(T_o)| 
    \end{equation}
\end{definition}
In other words, $\Delta P$ captures the incremental increase in the number of retweets after $T_o$, indicating the growth and spread of the cascade over time.

\section{Methodology}
\begin{figure*}[t]
    \centering
    \includegraphics[width=\textwidth]{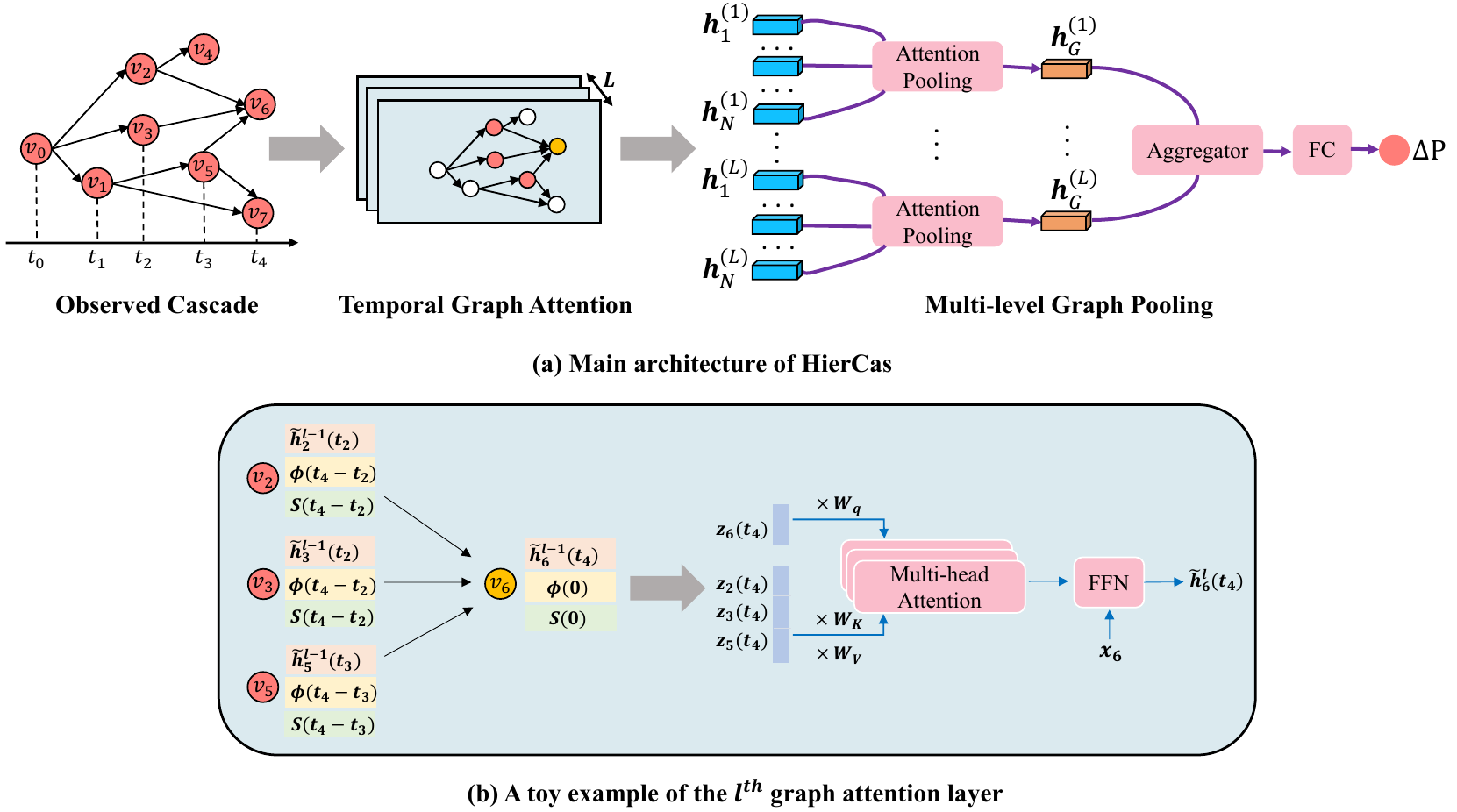}
    \caption{Detailed framework of HierCas. (a) is the main architecture. HierCas inputs the observed cascade graph, stacking several temporal graph attention layers to model temporal and structural information. Then the multi-level graph pooling module aggregates different localities learned by graph attention layers to make graph-level predictions. (b) takes $v_6$ as an example to show the node aggregation process in each layer, including the time-aware node embedding (time embedding \colorbox[RGB]{255,242,204}{$\Phi(\cdot)$} and size embedding \colorbox[RGB]{226,240,217}{$S(\cdot)$}) and graph attention mechanisms.}
    \label{fig:model}
\end{figure*}

In this section, we elaborate on the proposed model HierCas. The framework of HierCas takes the observed cascade as input and outputs the popularity increment, as shown in Figure \ref{fig:model}(a). HierCas consists of two main components: \textbf{Temporal Graph Attention} and \textbf{Multi-level Graph Pooling}. We will introduce each of the components in the rest of this section. 
\subsection{Temporal Graph Attention}
\subsubsection{Time-aware Node Embedding}
\ 
\newline
Considering each node represents each user, we encode user identity as the raw node representations. Specifically, for the cascade graph $\mathcal{G}\left(T_o\right)$, each node $v_i$ on the graph represents user $i$ who is involved in the forwarding process. The user identity can be represented as a one-hot vector, $\bm{u}_{i} \in \mathbb{R}^N$, where $N$ is the total number of users. All users share an embedding matrix $\bm{W}_{u} \in \mathbb{R}^{d_u \times N}$, where $d_u$ is the dimension of user embedding. The embedding matrix is used to transform the user identity into a dense representation vector denoted as
\begin{equation}
    \bm{x}_i = \bm{W}_u \bm{u}_i
\end{equation}
where $\bm{x}_{i} \in \mathbb{R}^{d_u}$. \\
In addition to user identity, our approach incorporates the relative time and the changes in cascade size in the timespan as the node features during the node aggregation process. Next, we introduce these two types of features separately.\par
\noindent
\textbf{Time embedding. }
In order to capture temporal dynamics and incorporate them into node representations, we introduce the concept of time embedding. The key to time embedding is how to capture the evolution of the cascade graph over time using retweet timestamps. Inspired by \cite{tgat_iclr20}, we aim to establish a function $\Phi: T \rightarrow \mathbb{R}^{d_t}$ that maps the time information to a $d_t$-dimensional vector space, enabling it to be embedded in the node representation. Specifically, consider two nodes $v_1, v_2$ on $\mathcal{G}(T_o)$ at time $t_1, t_2$ where $t_1 > t_2$, we define the mapping $\Phi$ as follow:
\begin{equation}
    \Phi(t_1 - t_2) = \sqrt{\frac{1}{d_t}} \left[ \alpha_1 cos(\omega_1 \Delta t),..., \alpha_{d_t} cos(\omega_{d_t}\Delta t) \right]
\end{equation}
where $\Delta t=t_1-t_2, \left\{ \omega_i, \alpha_i \mid i \in \left\{1,...,d_t\right\} \right\}$ are the trainable frequency factors and time-decay parameters, respectively. We utilize the time interval because, compared to absolute time, we can analyze the patterns of retweet frequency to capture the changing trend of the cascade graph. Additionally, we introduce the time decay effect by incorporating a decay factor into the time embedding. It is crucial for popularity prediction that the influence of retweets gradually diminishes as time passes. In contrast to the graph-level decay modeling approaches\cite{cao2017deephawkes,10.1145/2983323.2983812}, we apply the decay factor to the time embedding of individual nodes, enabling the model to capture the fine-grained node-to-node time decay effects. \\

\noindent
\textbf{Size embedding. }
In our task of predicting the cascade popularity increment, the macroscopic changes in the size of the cascade graph also serve as an important predictive feature. To capture this information, we embed the size changes into the node representation, combining it with the time embedding to capture both local short-term outbreaks and long-term trends in popularity. The size changes here indicate the number of new retweets generated in the specific timespan. As in the cascade graph, it represents the change in the number of edges. We adopt a learnable embedding layer that encodes the discrete size changes as dense features.\par 
Consider two nodes $v_1, v_2$ on $\mathcal{G}(T_o)$ at time $t_1, t_2$ where $t_1 > t_2$, we then define the function $S$:
\begin{equation}
    S(t_1 - t_2) = \bm{W}_s \delta(t_1, t_2)
\end{equation}
where $\bm{W}_s \in \mathbb{R}^{d_s \times N}$ is the embedding matrix, $d_s$ is the dimension of size embedding, and $\delta(t_1, t_2) = \mathbf{one\text{-}hot}(|\mathcal{E}(t_1)| - |\mathcal{E}(t_2)|), t_1 > t_2$ is the one-hot representation of the size changes between time points $t_1$ and $t_2$. By combining size change information and relative time information, the model is able to understand the dynamic pattern of the cascade graph, i.e, the growth rate of the cascade graph at different moments, so that it can predict the trend of future popularity increment.

\subsubsection{Graph Attention Mechanism}
\ 
\newline
HierCas utilizes the temporal graph attention layers to learn the dynamics of cascades. Differently from the original formulation of the temporal graph attention layer (first proposed in \cite{tgat_iclr20}), in our case, we incorporate more time-aware information, i.e. the size change of the cascade graph as such it allows the model to better capture the popularity trend. Fig. \ref{fig:model} (b) shows an example to illustrate how nodes are aggregated in each layer. Since the calculation process of each node is consistent, we specify it by node $v_0$ at time $t_0$. We use ${\tilde{\bm{h}}_i}^{l}(t_i)$ to denote the hidden representation for node $v_i$ at time $t_i$ from the $l^{th}$ layer. For $l=0$, the raw node feature is the user embedding $\bm{x}_i$. \par
For node $v_0$ at time $t_0$ in the $l^{th}$ layer, we consider its temporal neighborhood $\mathcal{N}(v_0;t_0)=\left\{v_1,...v_n\right\}$, where $t_1,...t_n < t_0$, such that $v_0$ can only interact with the nodes prior to it. For each neighbor $v_i$ in $\mathcal{N}(v_0;t_0)$, its representation  can be defined as:
$$
z_i(t_0) = \left[ {\tilde{\bm{h}}_i}^{l-1}(t_i) || \Phi(t_0-t_i) || S(t_0-t_i) \right]
$$
where $||$ is concatenation, $\Phi(t_0-t_i)$ is the time embedding that considers the timespan between $v_i$ and the target node $v_0$, and $S(t_0-t_i)$ is the size embedding with respect to the size change within that timespan. $z_i(t_0)$ takes into account the time-aware information as well as the hidden representation of the $(l-1)^{th}$ layer. In order to capture the connection between the target node and its neighbors, their representations will be fed into the attention layer. We denote the input of the attention layer as $\bm{Z}(t_0)$:
\begin{equation}
    \bm{Z}(t_0) = \left[z_0(t_0),...,z_i(t_0),...,z_n(t_0) \right]
\end{equation}
First, $\bm{Z}(t_0)$ is forwarded to three different linear projections to obtain 'query', 'key', and 'value':
\begin{equation}
\begin{aligned}
    \bm{q}(t_0) &= {\left[\bm{Z}(t_0)\right]}_0 \bm{W}_q \\
    \bm{K}(t_0) &= {\left[\bm{Z}(t_0)\right]}_{1:n} \bm{W}_K \\
    \bm{V}(t_0) &= {\left[\bm{Z}(t_0)\right]}_{1:n} \bm{W}_V
\end{aligned}
\end{equation}
    
where $\bm{W}_q, \bm{W}_K, \bm{W}_V \in \mathbb{R}^{(d_h+d_t+d_s) \times d_h}, d_h$ is the dimension of the hidden layer. These weight matrices are used to capture the connection of the temporal and structural information between layers. We then take the dot-product self-attention to compute hidden neighborhood representations $\bm{h}(t_0)$:
\begin{align}
    & \bm{h}(t_0) = Attn(\bm{q}(t_0), \bm{K}(t_0), \bm{V}(t_0)) \in \mathbb{R}^{d_h}, \\
    \label{eq:attn} & Attn(\bm{Q}, \bm{K}, \bm{V}) = softmax(\frac{\bm{Q}\bm{K}^T}{\sqrt{d_h}})\bm{V}
\end{align}
The attention weights $\{\alpha_i\}^N_{i=1}$ of the softmax output in Eq.\ref{eq:attn} is given by $\alpha_i=exp\left(\bm{Q}^T \bm{K}_i\right) / \left(\Sigma_q exp\left(\bm{Q}^T \bm{K}_q\right)\right)$. $\alpha_i$ reveals how node $v_i$ attends to the features of node $v_0$ with the topological structure. The self-attention defines a local temporal aggregation operator on the graph, therefore capturing the temporal interactions between nodes and the popularity trend over time.\par

To integrate the neighborhood representation with the target node's features, we concatenate the neighborhood representation with the raw feature vector of the target node, which corresponds to the user embedding $x_0$. This concatenated input is then fed into a feed-forward neural network to capture non-linear interactions between the features:
\begin{equation}
    \tilde{\bm{h}}_0^{(l)}(t_0) = \operatorname{ReLU}\left(\left[\bm{h}(t_0) \| \bm{x}_0\right] \bm{W}_0^{(l)}+\bm{b}_0^{(l)}\right) \bm{W}_1^{(l)}+\bm{b}_1^{(l)}    
\end{equation}
where $\bm{W}_0^{(l)} \in \mathbb{R}^{\left(d_h+d_u\right) \times d_h}, \bm{W}_1^{(l)} \in \mathbb{R}^{d_h \times d_h}, \text{and } \bm{b}_0^{(l)}, \bm{b}_1^{(l)} \in \mathbb{R}^{d_h}$ . \\ $\tilde{\bm{h}}_0^{(l)}(t_0) \in \mathbb{R}^{d_h}$ is the final output representing the time-aware node embedding at time $t_0$ for the target node $v_0$. \par

\subsection{Multi-level Graph Pooling}
The task of predicting cascade popularity involves performing graph-level regression, which requires the model to effectively incorporate both local information from individual nodes and global information from the entire graph. For this, we utilize a multi-level graph pooling method that hierarchically summarizes the information in the graph by progressively aggregating nodes at different levels of granularity. \par
For $\forall l \in \left\{1,..., L \right\}$, consider all node features $\left\{\tilde{\bm{h}}_0^{(l)}(t_0),...,\tilde{\bm{h}}_N^{(l)}(t_N) \right\}$ output by the $l^{th}$ layer, a dedicated attention-pooling function is used to compute layer-wise graph representation:
\begin{equation}\label{attpool}
\begin{aligned}
    \bm{h}_G^{(l)} &= AttPool^{(l)}\left( \left\{\tilde{\bm{h}}_0^{(l)}(t_0),...,\tilde{\bm{h}}_N^{(l)}(t_N) \right\} \right) \\
    & = \sum_{i=1}^{N} softmax\left( MLP^{(l)} ( \tilde{\bm{h}}_i^{(l)}(t_i))\right) \tilde{\bm{h}}_i^{(l)}(t_i)
\end{aligned}
\end{equation}
where $MLP^{(l)}$ is a two-layer neural network used to compute the attention score. This attention-pooling method can capture the contribution of different nodes to the graph representation in each layer.\par 
Since each layer is designed to capture distinct ranges of neighborhood information, we further employ a weighted sum operation to aggregate the layer-wise representations based on their importance:
\begin{equation}
    \bm{h}_G = \sum_{l=1}^L \omega^{(l)} \bm{h}_G^{(l)}
\end{equation}
where $\left\{ \omega^{(l)} \mid l \in \left\{ 1,...,L\right\} \right\}$ is a trainable weight vector. By utilizing the multi-level attention pooling method, our method can dynamically aggregate structural information from different localities in cascade graphs over time.\par

In the end, we feed the aggregated graph feature $\bm{h}_G$ into a fully connected layer to predict the incremental popularity $\Delta P_i$ by minimizing the mean square logarithmic error loss function:
\begin{equation}
    loss = \frac{1}{M} \sum_{i=1}^M \left(log \Delta \hat{P_i} - log \Delta P_i\right)^2
\end{equation}
where $M$ is the total number of messages, $\Delta \hat{P_i}$ is the ground-truth and $\Delta P_i$ is the predicted popularity for message $i$. 
\section{Experiments}

\subsection{Datasets}

To evaluate the effectiveness of the HierCas model, we apply it to two different popularity prediction scenarios, i.e., predicting the future size of retweets on Weibo and the citation count of papers from APS. These two scenarios have been extensively studied in previous work \cite{chen2019cascn,cao2017deephawkes,CasHAN,casseqgcn,casflow,wang2020predicting} and represent two typical cascade modes, i.e. weibo is short-term, dramatic cascading evolution while APS is long-term and smooth, which provides a broad basis for comparing the proposed model with state-of-the-art methods. \par
Our experiments were conducted using the same dataset settings as in previous studies\cite{casflow,mucas}. For Weibo dataset, we set the observation time window $T_o$ to 0.5h and 1h, while for the APS dataset, we set $T_o$ to 3 and 5 years. The prediction time $T_p$ is 24 hours for Weibo dataset and 20 years for APS dataset. We filter out cascades with observed size $|\mathcal{E}(t_o)| < 10$ to focus on significant cascades. For larger cascades with $|\mathcal{E}(t_o)| > 100$, we only utilize the first 100 retweets as model input to ensure computational efficiency. We randomly split each dataset into a training set (70\%), a validation set (15\%), and a testing set (15\%). Descriptive statistics of two datasets are shown in Table \ref{tab:stats}. 

\begin{table}
    \caption{Descriptive statistics of two datasets.}
    \label{tab:stats}
    \centering
    \begin{tabular}{lcccc}
        \hline 
        \textbf{Datasets} & \multicolumn{2}{c}{ \textbf{Sina Weibo} } & \multicolumn{2}{c}{ \textbf{APS} } \\
        \hline
        $\#$ All Nodes & \multicolumn{2}{c}{6,738,040} & \multicolumn{2}{c}{690,216} \\
        $\#$ All Edges & \multicolumn{2}{c}{20,642,990} & \multicolumn{2}{c}{9,370,286} \\
        \hline
        \multicolumn{5}{c}{ \textit{Number of cascades} }\\
        $\mathrm{T}_o$ & $0.5 \text{ hour}$ & $1 \text{ hour}$  & $3 \text{ years}$ & $5 \text{ years}$ \\
        train & 19,472 & 24,646 & 19,270 & 33,582  \\
        val & 4,173 & 5,279 & 4,129 & 7,196 \\
        test & 4,172 & 5,278 & 4,130 & 7,197 \\
        \hline
        \multicolumn{5}{c}{ \textit{Average number of hops} } \\
        $\mathrm{T}_o$ & $0.5 \text{ hour}$ & $1 \text{ hour}$  & $3 \text{ years}$ & $5 \text{ years}$ \\
        train & 2.17 & 2.17 & 3.89 & 3.91  \\
        val & 2.15 & 2.17 & 3.87 & 3.92 \\
        test & 2.14 & 2.17 & 3.88 & 3.91 \\
        \hline
        \multicolumn{5}{c}{ \textit{Average growth popularity} } \\
        $\mathrm{T}_o$ & $0.5 \text{ hour}$ & $1 \text{ hour}$  & $3 \text{ years}$ & $5 \text{ years}$ \\
        train & 280.6 & 242.7 & 52.7 & 41.1  \\
        val & 351.7 & 272.7  & 53.4 & 42.4 \\
        test & 432.9 & 299.6  & 50.0 & 41.7 \\
        
        \hline
    \end{tabular}
    
\end{table}
\subsection{Experimental Setups}

\begin{table*}[t]
    \caption{Overall comparison between baselines and HierCas on two datasets with different observation times. The best results are in \textbf{bold} and the best baseline results are \underline{underlined}. The improvements achieved by HierCas compared to the best baseline are also provided. A paired t-test is performed and $*$ indicates a statistical significance $p < 0.001$ as compared to the best baseline.}
    \label{tab:comparison}
    \centering
    \resizebox{\linewidth}{!}{
    \begin{tabular}{lcccccccccccc}
        \toprule
        \multirow{3}{*}{\textbf{Model}} &  \multicolumn{6}{c}{\textbf{Weibo}} & \multicolumn{6}{c}{\textbf{APS}}\\
        \cmidrule(r){2-7} \cmidrule{8-13}
        & \multicolumn{3}{c}{\textbf{0.5 Hour}} &  \multicolumn{3}{c}{\textbf{1 Hour}} & \multicolumn{3}{c}{\textbf{3 Years}} & \multicolumn{3}{c}{\textbf{5 Years}}  \\
        \cmidrule(r){2-4} \cmidrule(r){5-7} \cmidrule(r){8-10} \cmidrule{11-13} 
        & \textbf{MSLE} & \textbf{SMAPE} & $\bm{R^2}$ & \textbf{MSLE} & \textbf{SMAPE} & $\bm{R^2}$ & \textbf{MSLE} & \textbf{SMAPE} & $\bm{R^2}$ & \textbf{MSLE} & \textbf{SMAPE} & $\bm{R^2}$ \\
        \midrule
        DeepCas\cite{li2017deepcas} & 2.911 & 0.327 & 0.354 & 2.743 & 0.355 & 0.451 & 1.978 & 0.349 & 0.198 & 1.932 & 0.357 & 0.299 \\
        DeepHawkes\cite{cao2017deephawkes} & 2.891 & 0.321 & 0.379 & 2.632 & 0.352 & 0.468 & 1.823 & 0.332 & 0.228 & 1.564 & 0.337 & 0.351 \\
        CasHAN\cite{CasHAN} & 2.212 & 0.231 & 0.615 & 2.006 & 0.235 & 0.631 & 1.413 & 0.254 & 0.405 & 1.413 & 0.281 & 0.452 \\
        \midrule
        CasCN\cite{chen2019cascn} & 2.804 & 0.311 & 0.381 & 2.601 & 0.350 & 0.468 & 1.807 & 0.271 & 0.232 & 1.543 & 0.336 & 0.354 \\
        CasGCN\cite{xu2020casgcn} & 2.621 & 0.289 & 0.453 & 2.523 & 0.322 & 0.567 & 1.675 & 0.266 & 0.287 & 1.521 & 0.321 & 0.376 \\
        CasSeqGCN\cite{casseqgcn} & 2.432 & 0.256 & 0.589 & 2.341 & 0.297 & 0.604 & 1.523 & 0.261 & 0.343 & 1.487 & 0.311 & 0.386 \\
        MUCas\cite{mucas} & \underline{2.140} & 0.277 & 0.612 & \underline{1.978} & 0.311 & 0.629 & 1.493 & 0.274 & 0.344 & \underline{1.399} & 0.335 & 0.388 \\ 
        CasFlow\cite{casflow} & 2.171 & \underline{0.202} &  \underline{0.618} & 1.979 & \underline{0.223} & \underline{0.649} & \underline{1.387} & \underline{0.233} & \underline{0.462} & 1.422 & \underline{0.269} & \underline{0.498} \\

        \midrule
        HierCas-noTime & 2.183 & 0.195 & 0.568 & 2.132 & 0.221 & 0.584 & 1.066 & 0.189 & 0.435 & 0.883 & 0.196 & 0.493 \\

        HierCas-noSize & 2.399 & 0.202 & 0.526 & 2.259 & 0.219 & 0.560 & 1.080 & 0.189 & 0.427 & 0.913 & 0.197 & 0.476 \\
        
        HierCas-meanAgg & 2.386 & 0.203 & 0.528 & 2.273 & 0.227 & 0.556 & 1.035 & 0.185 & 0.445 & 0.881 & 0.196 & 0.485 \\
        
        HierCas-noMulti & 1.952 & 0.188 & 0.614 & 1.779 & 0.201 & 0.653 & 0.964 & 0.179 & 0.487 & 0.806 & 0.189 & 0.538 \\
        
        HierCas (Ours) & \bm{$1.817^*$} & \bm{$0.186^*$} & \bm{$0.633^*$} & \bm{$1.695^*$} & \bm{$0.194^*$} & \bm{$0.672^*$} & \bm{$0.944^*$} & \bm{$0.177^*$} & \bm{$0.497^*$} & \bm{$0.769^*$} & \bm{$0.187^*$} & \bm{$0.551^*$} \\
        (Improvements) & $\uparrow 15.1\%$ & $\uparrow 7.9\%$ & $\uparrow 2.4\%$ & $\uparrow 14.3\%$ & $\uparrow 13.0\%$ & $\uparrow 3.5\%$ & $\uparrow 31.9\%$ & $\uparrow 24.0\%$ & $\uparrow 7.6\%$ & $\uparrow 45.9\%$ & $\uparrow 30.4\%$ & $\uparrow 10.6\%$ \\
        \bottomrule
    \end{tabular}}
    
\end{table*}

\subsubsection{Evaluation metrics}
Following previous work \cite{casflow, mucas}, we use mean square logarithmic error (MSLE), symmetric mean absolute percentage error (SMAPE), and the coefficient of determination ($R^2$) for prediction performance evaluation. The formulation of all evaluation metrics is defined as: 

$$
\begin{aligned}
    \operatorname{MSLE} & =\frac{1}{M} \sum_{i=1}^M\left(\log \Delta P_i-\log \Delta \hat{P_i}\right)^2, \\
    \mathrm{SMAPE} & =\frac{1}{M} \sum_{i=1}^M \frac{\left|\log \Delta P_i-\log \Delta \hat{P_i}\right|}{\left(\left|\log \Delta P_i\right|+\left|\log \Delta \hat{P_i}\right|\right) / 2}, \\
    R^2 & =1-\frac{\sum_{i=1}^M\left(\log \Delta P_i-\log \Delta \hat{P_i}\right)^2}{\sum_{i=1}^M\left(\log \Delta \hat{P_i}-\frac{1}{M} \sum_{i=1}^M \log \Delta \hat{P_i}\right)^2},
\end{aligned}
$$

where $M$ is the total number of messages, $\Delta P_i$ is the predicted incremental popularity of message $i$, and $\hat{P}_i$ is the ground-truth. Note that, the value of MSLE and SMAPE the lower the better, in contrast, the value of $R^2$ the higher the better.

\subsubsection{Baselines and Parameter Settings}
To validate HierCas's performance in cascade prediction, we select three RNN-based and five GNN-based methods for comparison.
For DeepCas, DeepHawkes, and CasCN, the node embedding size is set to 50, which is consistent with \cite{casflow,mucas}. For CasHAN, the balance factor $\beta$ is set to $0.6$. For MUCas, the embedding size of positional embedding is set to 50. The max-order $K$ and iteration number $\tau$ are chosen to 3. For all baselines, the batch size is 64, the learning rate is set to $3 \times 10^{-5}$, and all the other hyper-parameters are set to the same values as used in the original papers.\par
As for our model HierCas, the batch size and learning rate are the same as the baselines. We use AdamW as the optimizer. The number of graph attention layers $L$ is set to $2$. For each node $v_i$ in the graph, we uniformly sample its $20$ neighbors from $\mathcal{N}(v_i)$ for node aggregation. The dimension of node embedding $d_u, d_t, d_s$ and hidden layer $d_h$ are set to $64$. The number of attention heads is 4. In terms of time embedding, we use $seconds$ for Weibo dataset and $days$ for APS dataset to calculate the relative time interval, taking into account the difference between tweets and research papers. The experiments are conducted on one NVIDIA GeForce GTX 3090 with 24GB memory.
\subsection{Overall Performance}
We compare the prediction performance of HierCas with eight strong baselines on two datasets. Table \ref{tab:comparison} shows the overall comparisons, from which we have the following observations: 
(1) The proposed HierCas model outperforms all the baselines in both scenarios under three evaluation metrics. For the prediction error MSLE, HierCas surpasses the best baseline by $14.7\%$ and $38.9\%$ on average for the two datasets, respectively, which is a significant improvement. HierCas exhibits a notable performance improvement on the APS dataset compared to the Weibo dataset. This discrepancy can be attributed to the dataset characteristics. For Weibo dataset, the presence of shorter propagation paths and rapid popularity growth adds more complexity to the prediction task.

(2) Among the RNN-based models, CasHAN uses node-level and sequence-level attention to enhance node sequence representation, but its performance is still restricted to their overlook of crucial structural information. The GNN-based methods, like MUCas and CasFlow, although adopting novel and complex graph learning methods to extract structural features, still fall short in the sampling method and separate modeling of temporal and structural information. CasGCN utilizes a static graph approach to modeling that ignores cascade graph changes over time. Our HierCas employs dynamic graph learning on the entire cascade graph, incorporating time-aware information, ensuring its optimal performance.


\begin{figure*}[t]
    \centering  
    \subfigure[Weibo - 0.5 hour]{
    \label{weibo_0.5h}
    \includegraphics[width=0.48\textwidth]{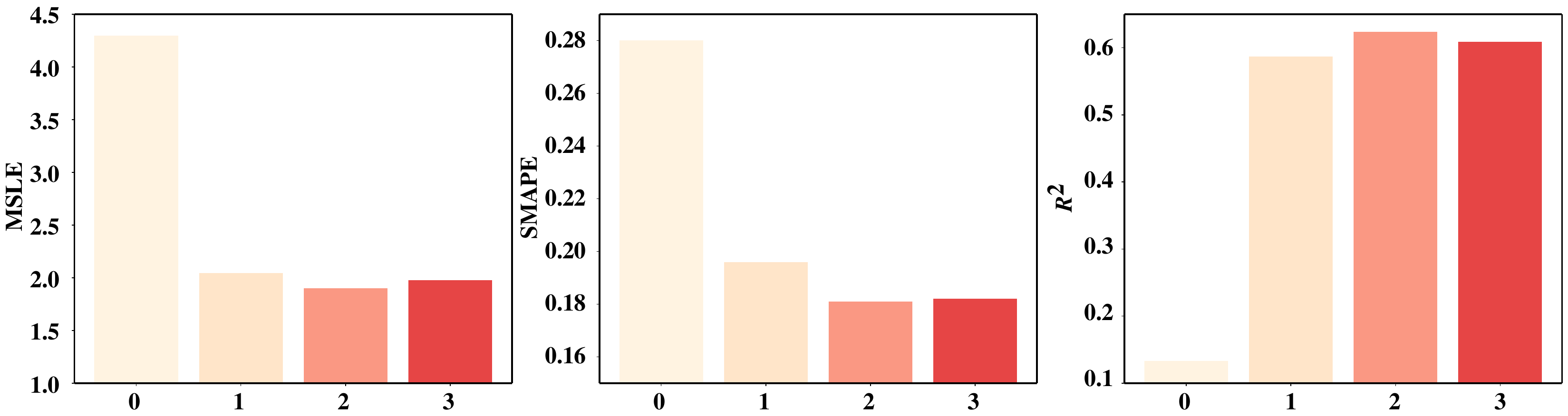}}
    \subfigure[Weibo - 1 hour]{
    \label{weibo_1h}
    \includegraphics[width=0.48\textwidth]{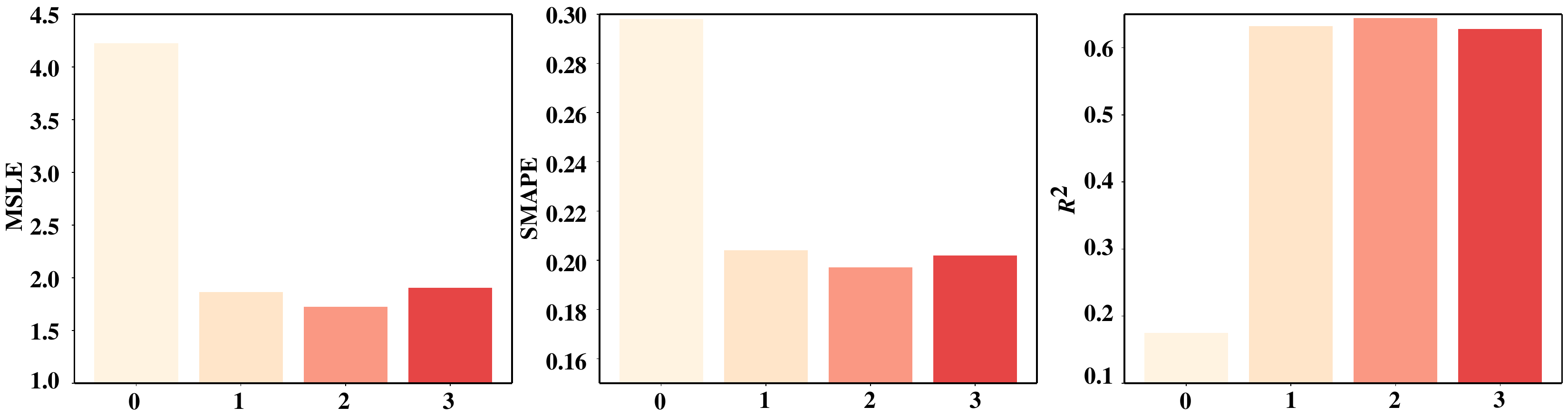}}
    \subfigure[APS - 3 years]{
    \label{aps_3y}
    \includegraphics[width=0.48\textwidth]{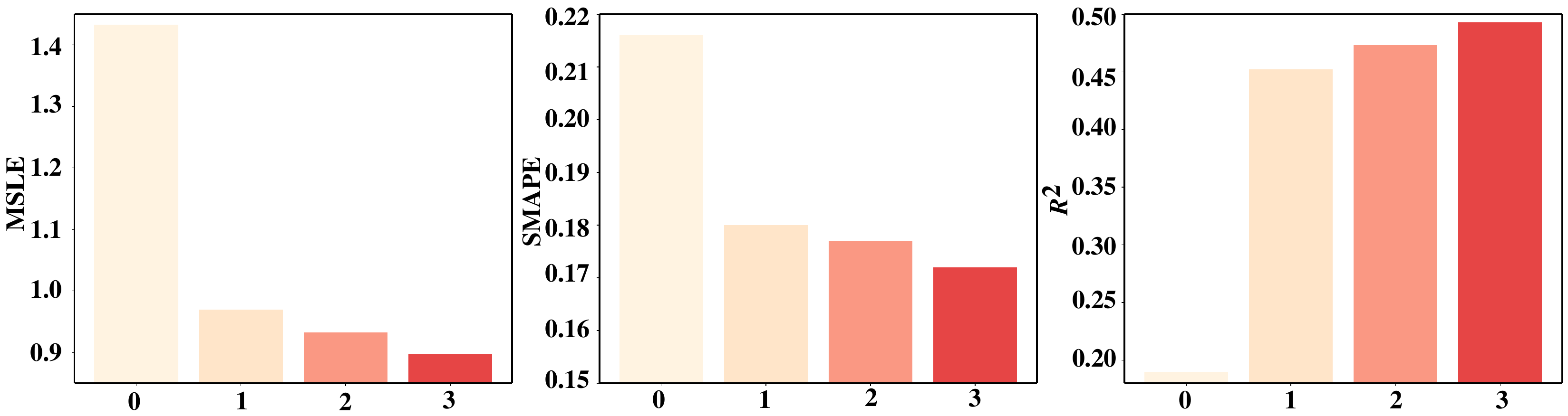}}
    \subfigure[APS - 5 years]{
    \label{aps_5y}
    \includegraphics[width=0.48\textwidth]{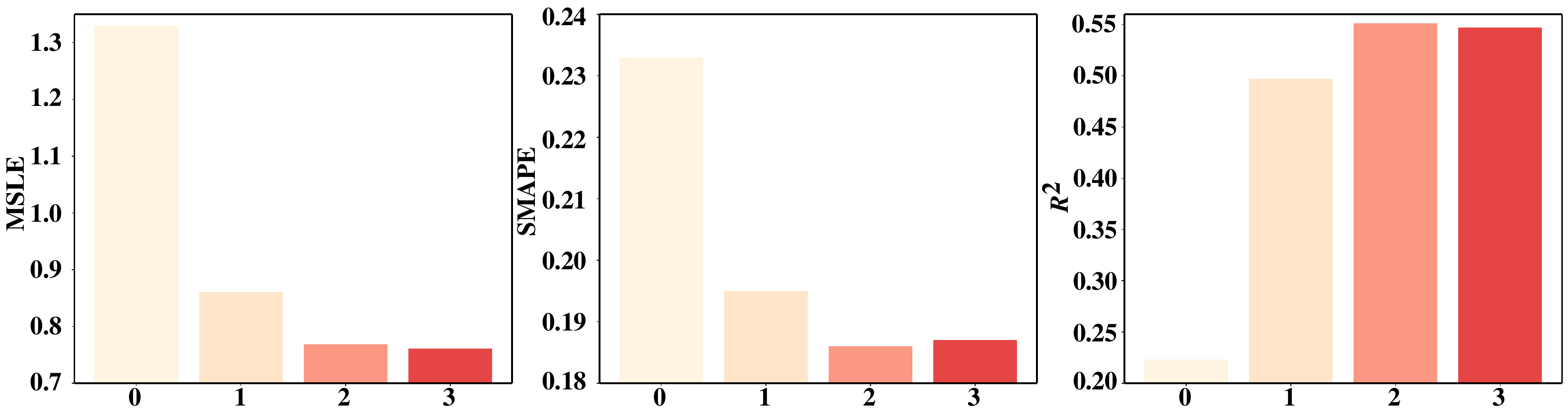}}
    \caption{Impact of the number of graph attention layers.}
    \label{fig:layer}
\end{figure*}

\subsection{Ablation Study}
We conduct ablation experiments to investigate the contributions of different components in HierCas. The results are shown in Table \ref{tab:comparison}. Regarding the different variant models in the ablation study, we draw the following conclusions: 

\noindent\textbf{HierCas-noTime:} By removing the temporal feature $\Phi$ from the node embedding module, the model's performance slightly deteriorates in terms of all metrics, suggesting that incorporating temporal information is essential for capturing the dynamics of cascades and improving prediction accuracy.

\noindent\textbf{HierCas-noSize:} Removing the size feature $S$ from the node embedding module results in a significant decrease in performance across all metrics. This demonstrates that considering the size changes of cascades can capture the trend in popularity, which is crucial for cascade size prediction.

\noindent\textbf{HierCas-meanAgg:} Replacing the graph attention mechanism with simple averaging in the neighbor aggregation process leads to inferior performance compared to the full HierCas model. This indicates that the attention mechanism effectively captures relevant information from the neighbors, contributing to better node representation.

\noindent\textbf{HierCas-noMulti:} By only using the pooled graph-level feature $h_G^{(l)}$ of the last graph attention layer for prediction, the model's performance drops slightly. This highlights the importance of multi-level pooling for aggregating different temporal and structural localities from different layers.

\begin{figure}[t]
    \centering  
    \subfigure[Weibo - 0.5 hour]{
    \label{weibo_0.5h}
    \includegraphics[width=0.23\textwidth]{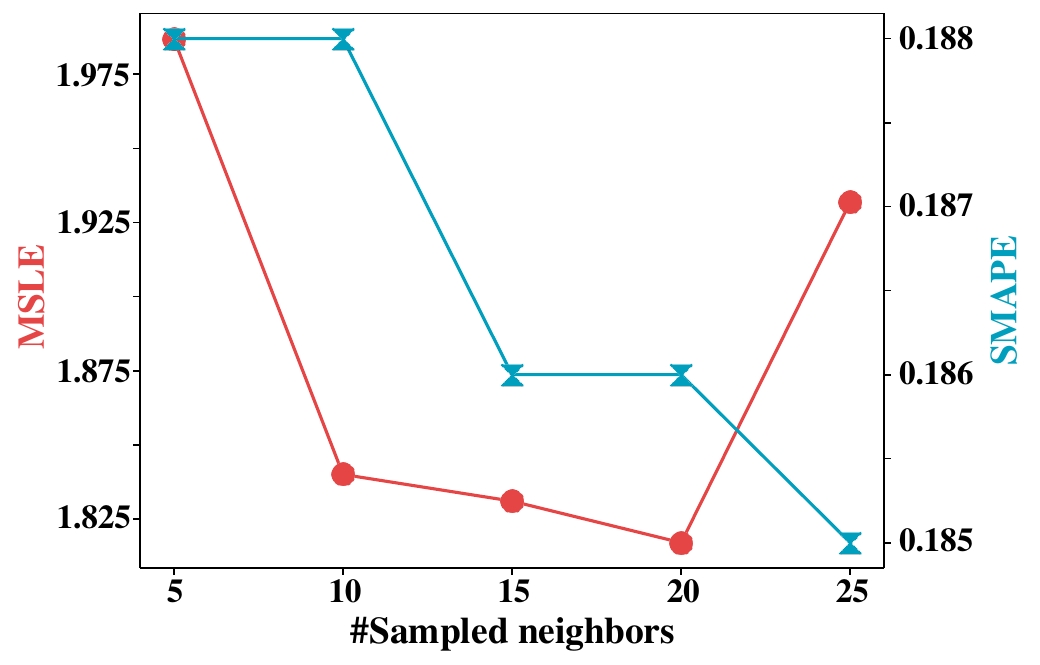}}
    \subfigure[Weibo - 1 hour]{
    \label{weibo_1h}
    \includegraphics[width=0.23\textwidth]{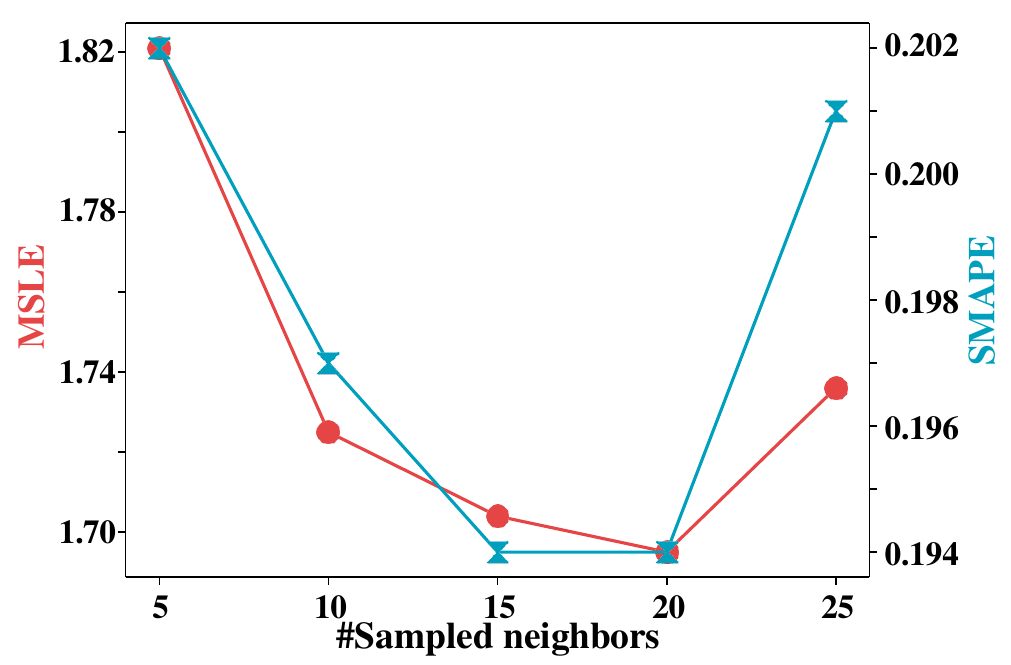}}
    \subfigure[APS - 3 years]{
    \label{aps_3y}
    \includegraphics[width=0.23\textwidth]{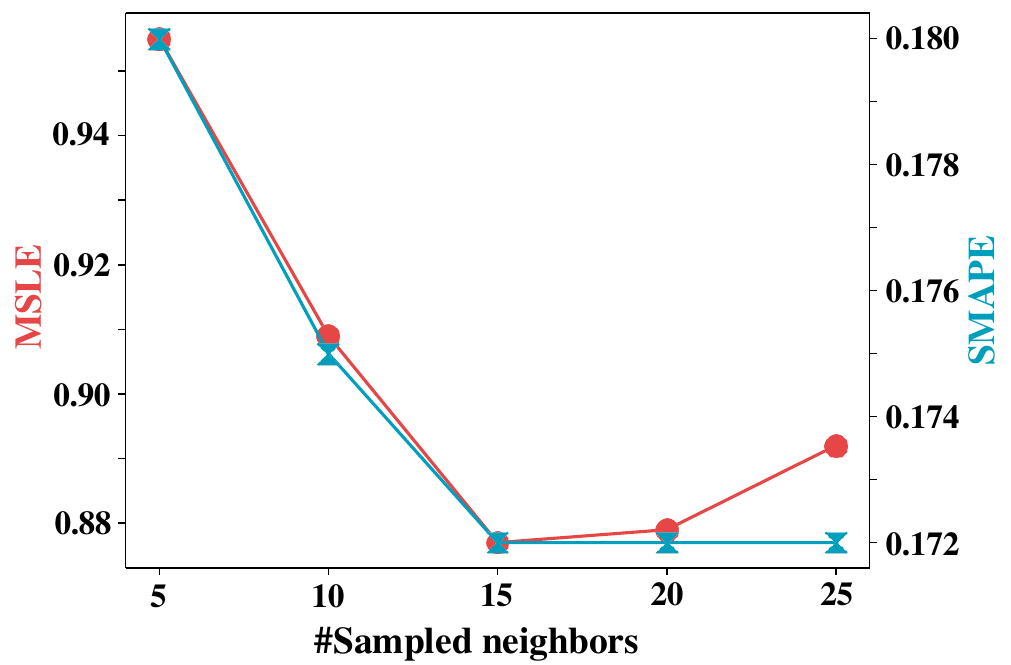}}
    \subfigure[APS - 5 years]{
    \label{aps_5y}
    \includegraphics[width=0.23\textwidth]{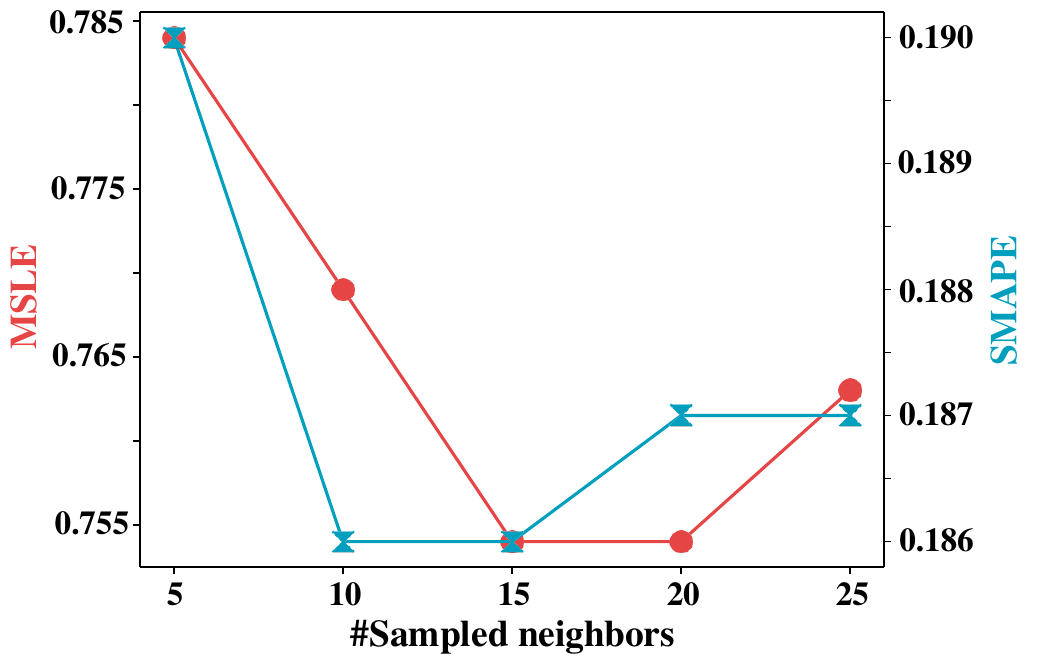}}
    \caption{Impact of the number of neighbors.}
    \label{fig:neighbors}
\end{figure}

\subsection{Parameter Analysis}
\subsubsection{The number of graph attention layers}
We analyze the important parameter, the number of graph attention layers $L$, on both datasets. We vary $L$ from $0$ to $3$ to observe its effect on prediction results. Each additional layer represents the aggregation of one more hop of neighbors for each node, while the $0$-layer configuration indicates no structural modeling. On both datasets, we observe that increasing the number of graph attention layers from $0$ to $2$ leads to consistent improvements in performance metrics. On the Weibo dataset, we notice a slight decrease in the results as $L$ increased from 2 to 3. Conversely, on the APS dataset, there was still an improvement in the prediction error, specifically in terms of MLSE. We can infer that the optimal number of graph attention layers may depend on the average number of hops in the dataset. Datasets with longer average hops, like APS, may benefit from a larger number of layers to capture the long-range dependencies.
\subsubsection{The number of neighbors}
We further investigate the impact of the number of sampled neighbors on the prediction performance. We vary the number of samples in $\{5, 10, 15, 20, 25\}$ and analyze two evaluation metrics, MSLE represented by the red line and SMAPE represented by the blue line.
As illustrated in Fig. \ref{fig:neighbors}, we observe that increasing the number of sampled neighbors positively affects the prediction performance. The model benefits from aggregating information from a larger neighborhood, allowing it to capture more contextual cues and improve accuracy. However, we notice that after reaching a certain threshold (around 20 in our experiments), further increasing the number of samples leads to a decrease in performance. This may be due to the inclusion of noisy information from less relevant nodes, which negatively impacts the model's ability to make accurate predictions.

\subsection{Model Interpretability}
\begin{figure}
    \centering  
    \begin{minipage}{0.24\textwidth}
        \centering
        \subfigure[Weibo-0.5 hour]{
        \label{vis1}
        \includegraphics[width=\textwidth]{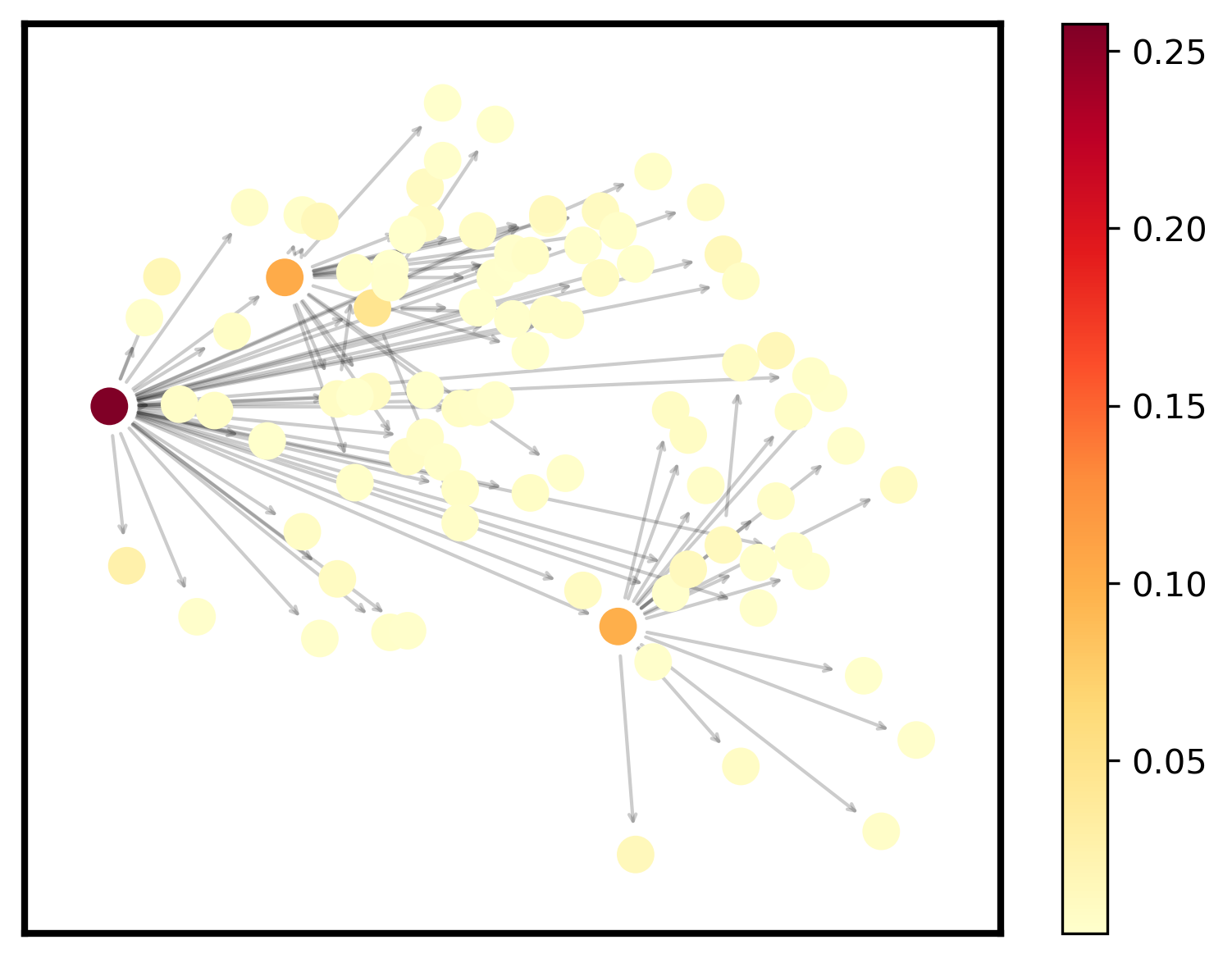}
        }
    \end{minipage}
    \begin{minipage}{0.24\textwidth}
        \centering
        \subfigure[APS-3 years]{
        \label{vis2}
        \includegraphics[width=\textwidth]{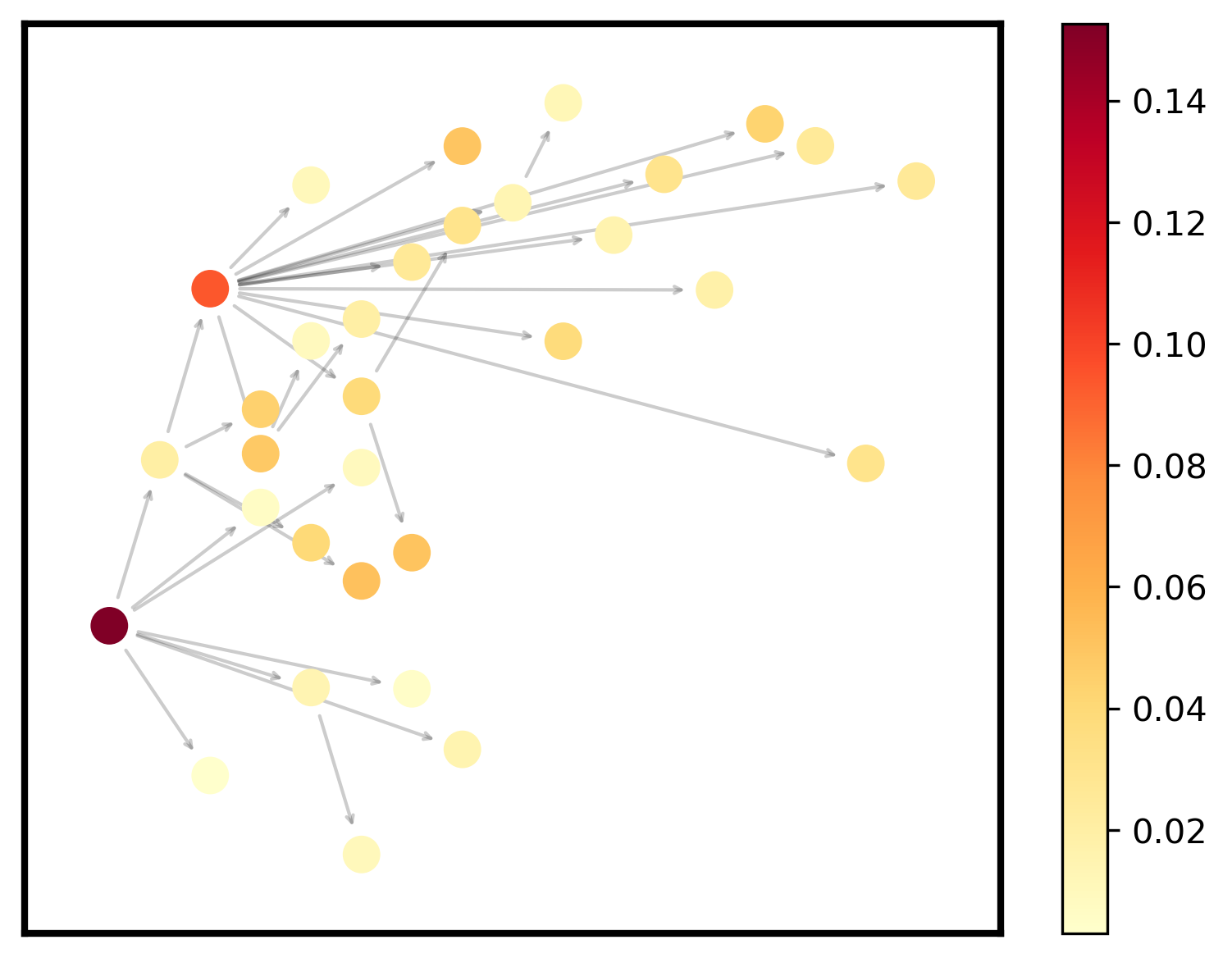}
        }
    \end{minipage}
    \caption{Visualization of nodes' contributions.}
    \label{fig:vis}
\end{figure}
Fig. \ref{fig:vis} visualizes two representative cascade graphs and attention coefficients of the $L^{th}$ pooling layer computed by softmax in Eq. \ref{attpool}, which can give us an insight into understanding how complex cascades emerge. The color of nodes indicates their contributions to the cascade. It can be seen that the cascade in the Weibo dataset evolves faster and at a larger scale. And for both datasets, the root node plays an important role. Some hub nodes also contribute a lot. Conversely, other nodes have little effect on the cascade evolution. 
\section{Conclusion}
We have presented a novel framework, named HierCas, for cascade popularity prediction. Through dynamic cascade modeling with time-aware node embedding, graph attention mechanisms, and hierarchical pooling structures, HierCas effectively learns the popularity trend implicit in the complex cascade. Experimental results on two real-world datasets from different scenarios demonstrate the superior performance of HierCas. 
In future work, we aim to incorporate temporal point processes such as the Hawkes process in neural networks to further reveal the intrinsic mechanisms behind the emergence of complex cascades.

\section*{Acknowledgements}
This research was sponsored by National Natural Science Foundation of China with grant number 62072048.

\bibliographystyle{IEEEtran}
\bibliography{conference_101719}

\begin{thebibliography}{10}
\providecommand{\url}[1]{#1}
\csname url@samestyle\endcsname
\providecommand{\newblock}{\relax}
\providecommand{\bibinfo}[2]{#2}
\providecommand{\BIBentrySTDinterwordspacing}{\spaceskip=0pt\relax}
\providecommand{\BIBentryALTinterwordstretchfactor}{4}
\providecommand{\BIBentryALTinterwordspacing}{\spaceskip=\fontdimen2\font plus
\BIBentryALTinterwordstretchfactor\fontdimen3\font minus \fontdimen4\font\relax}
\providecommand{\BIBforeignlanguage}[2]{{%
\expandafter\ifx\csname l@#1\endcsname\relax
\typeout{** WARNING: IEEEtran.bst: No hyphenation pattern has been}%
\typeout{** loaded for the language `#1'. Using the pattern for}%
\typeout{** the default language instead.}%
\else
\language=\csname l@#1\endcsname
\fi
#2}}
\providecommand{\BIBdecl}{\relax}
\BIBdecl

\bibitem{dow2013anatomy}
P.~A. Dow, L.~Adamic, and A.~Friggeri, ``The anatomy of large facebook cascades,'' in \emph{Proceedings of the International AAAI Conference on Web and Social Media}, 2013, pp. 145--154.

\bibitem{AlFalahi}
K.~AlFalahi, Y.~Atif, and A.~Abraham, ``Models of influence in online social networks,'' \emph{International Journal of Intelligent Systems}, pp. 161--183, 2014.

\bibitem{doi:10.1126/science.aao2998}
D.~M.~J. Lazer, M.~A. Baum, Y.~Benkler, A.~J. Berinsky, K.~M. Greenhill, F.~Menczer, M.~J. Metzger, B.~Nyhan, G.~Pennycook, D.~Rothschild, M.~Schudson, S.~A. Sloman, C.~R. Sunstein, E.~A. Thorson, D.~J. Watts, and J.~L. Zittrain, ``The science of fake news,'' \emph{Science}, pp. 1094--1096, 2018.

\bibitem{9078874}
N.~Ta, K.~Li, Y.~Yang, F.~Jiao, Z.~Tang, and G.~Li, ``Evaluating public anxiety for topic-based communities in social networks,'' \emph{IEEE Transactions on Knowledge and Data Engineering}, pp. 1191--1205, 2022.

\bibitem{deepinf}
J.~Qiu, J.~Tang, H.~Ma, Y.~Dong, K.~Wang, and J.~Tang, ``Deepinf: Social influence prediction with deep learning,'' in \emph{Proceedings of the 24th ACM SIGKDD International Conference on Knowledge Discovery \& Data Mining}, 2018, p. 2110–2119.

\bibitem{10.5555/3367471.3367602}
C.~Yang, J.~Tang, M.~Sun, G.~Cui, and Z.~Liu, ``Multi-scale information diffusion prediction with reinforced recurrent networks,'' in \emph{Proceedings of the 28th International Joint Conference on Artificial Intelligence}, 2019, p. 4033–4039.

\bibitem{li2017deepcas}
C.~Li, J.~Ma, X.~Guo, and Q.~Mei, ``Deepcas: An end-to-end predictor of information cascades,'' in \emph{Proceedings of the 26th International Conference on World Wide Web}, 2017, p. 577–586.

\bibitem{cao2017deephawkes}
Q.~Cao, H.~Shen, K.~Cen, W.~Ouyang, and X.~Cheng, ``Deephawkes: Bridging the gap between prediction and understanding of information cascades,'' in \emph{Proceedings of the 2017 ACM on Conference on Information and Knowledge Management}, 2017, p. 1149–1158.

\bibitem{chen2019cascn}
X.~Chen, F.~Zhou, K.~Zhang, G.~Trajcevski, T.~Zhong, and F.~Zhang, ``Information diffusion prediction via recurrent cascades convolution,'' in \emph{2019 IEEE 35th international conference on data engineering (ICDE)}.\hskip 1em plus 0.5em minus 0.4em\relax IEEE, 2019, pp. 770--781.

\bibitem{PPOC}
G.~Szabo and B.~A. Huberman, ``Predicting the popularity of online content,'' \emph{Commun. ACM}, p. 80–88, 2010.

\bibitem{10.1145/2433396.2433443}
H.~Pinto, J.~M. Almeida, and M.~A. Gon\c{c}alves, ``Using early view patterns to predict the popularity of youtube videos,'' in \emph{Proceedings of the Sixth ACM International Conference on Web Search and Data Mining}, 2013, p. 365–374.

\bibitem{10.1145/2566486.2567997}
J.~Cheng, L.~Adamic, P.~A. Dow, J.~M. Kleinberg, and J.~Leskovec, ``Can cascades be predicted?'' in \emph{Proceedings of the 23rd International Conference on World Wide Web}, 2014, p. 925–936.

\bibitem{Shulman2016PredictabilityOP}
B.~Shulman, A.~Sharma, and D.~Cosley, ``Predictability of popularity: Gaps between prediction and understanding,'' in \emph{International Conference on Web and Social Media}, 2016.

\bibitem{10.1145/2806416.2806505}
W.~Ding, Y.~Shang, L.~Guo, X.~Hu, R.~Yan, and T.~He, ``Video popularity prediction by sentiment propagation via implicit network,'' in \emph{Proceedings of the 24th ACM International on Conference on Information and Knowledge Management}, 2015, p. 1621–1630.

\bibitem{10.1145/2661829.2662055}
B.~Chang, H.~Zhu, Y.~Ge, E.~Chen, H.~Xiong, and C.~Tan, ``Predicting the popularity of online serials with autoregressive models,'' in \emph{Proceedings of the 23rd ACM International Conference on Conference on Information and Knowledge Management}, 2014, p. 1339–1348.

\bibitem{10.1145/2623330.2623732}
B.~Perozzi, R.~Al-Rfou, and S.~Skiena, ``Deepwalk: Online learning of social representations,'' in \emph{Proceedings of the 20th ACM SIGKDD International Conference on Knowledge Discovery and Data Mining}, 2014, p. 701–710.

\bibitem{wang2017topological}
J.~Wang, V.~W. Zheng, Z.~Liu, and K.~C.-C. Chang, ``Topological recurrent neural network for diffusion prediction,'' in \emph{2017 IEEE international conference on data mining (ICDM)}, 2017, pp. 475--484.

\bibitem{10.5555/3172077.3172305}
Y.~Wang, H.~Shen, S.~Liu, J.~Gao, and X.~Cheng, ``Cascade dynamics modeling with attention-based recurrent neural network,'' in \emph{Proceedings of the 26th International Joint Conference on Artificial Intelligence}, 2017, p. 2985–2991.

\bibitem{tang2021fully}
X.~Tang, D.~Liao, W.~Huang, J.~Xu, L.~Zhu, and M.~Shen, ``Fully exploiting cascade graphs for real-time forwarding prediction,'' in \emph{Proceedings of the AAAI Conference on Artificial Intelligence}, 2021.

\bibitem{CasHAN}
C.~Zhong, F.~Xiong, S.~Pan, L.~Wang, and X.~Xiong, ``Hierarchical attention neural network for information cascade prediction,'' \emph{Information Sciences}, pp. 1109--1127, 2023.

\bibitem{wang2020predicting}
M.~Wang and K.~Li, ``Predicting information diffusion cascades using graph attention networks,'' in \emph{Neural Information Processing: 27th International Conference, ICONIP 2020, Bangkok, Thailand, November 18--22, 2020, Proceedings, Part IV 27}, 2020, pp. 104--112.

\bibitem{casseqgcn}
Y.~Wang, X.~Wang, Y.~Ran, R.~Michalski, and T.~Jia, ``Casseqgcn: Combining network structure and temporal sequence to predict information cascades,'' \emph{Expert Systems with Applications}, 2022.

\bibitem{mucas}
X.~Chen, F.~Zhang, F.~Zhou, and M.~Bonsangue, ``Multi‐scale graph capsule with influence attention for information cascades prediction,'' \emph{Int. J. Intell. Syst.}, p. 2584–2611, 2022.

\bibitem{casflow}
X.~Xu, F.~Zhou, K.~Zhang, S.~Liu, and G.~Trajcevski, ``Casflow: Exploring hierarchical structures and propagation uncertainty for cascade prediction,'' \emph{IEEE Transactions on Knowledge and Data Engineering}, pp. 3484--3499, 2023.

\bibitem{huang2019cascade2vec}
Z.~Huang, Z.~Wang, and R.~Zhang, ``Cascade2vec: Learning dynamic cascade representation by recurrent graph neural networks,'' \emph{IEEE Access}, pp. 144\,800--144\,812, 2019.

\bibitem{zhou2020variational}
F.~Zhou, X.~Xu, K.~Zhang, G.~Trajcevski, and T.~Zhong, ``Variational information diffusion for probabilistic cascades prediction,'' in \emph{IEEE INFOCOM 2020-IEEE Conference on Computer Communications}.\hskip 1em plus 0.5em minus 0.4em\relax IEEE, 2020, pp. 1618--1627.

\bibitem{xu2020casgcn}
Z.~Xu, M.~Qian, X.~Huang, and J.~Meng, ``Casgcn: Predicting future cascade growth based on information diffusion graph,'' 2020.

\bibitem{lu2023continuoustime}
X.~Lu, S.~Ji, L.~Yu, L.~Sun, B.~Du, and T.~Zhu, ``Continuous-time graph learning for cascade popularity prediction,'' 2023.

\bibitem{10.1145/3580305.3599281}
S.~Ji, X.~Lu, M.~Liu, L.~Sun, C.~Liu, B.~Du, and H.~Xiong, ``Community-based dynamic graph learning for popularity prediction,'' in \emph{Proceedings of the 29th ACM SIGKDD Conference on Knowledge Discovery and Data Mining}, 2023, p. 930–940.

\bibitem{tgat_iclr20}
D.~Xu, C.~Ruan, E.~Korpeoglu, S.~Kumar, and K.~Achan, ``Inductive representation learning on temporal graphs,'' in \emph{International Conference on Learning Representations (ICLR)}, 2020.

\bibitem{10.1145/2983323.2983812}
S.~Mishra, M.-A. Rizoiu, and L.~Xie, ``Feature driven and point process approaches for popularity prediction,'' in \emph{Proceedings of the 25th ACM International on Conference on Information and Knowledge Management}, 2016, p. 1069–1078.

\end{thebibliography}

\end{document}